\documentclass[twocolumn,aps,prd,superscriptaddress,floatfix]{revtex4}
\usepackage{graphicx}
\usepackage{dcolumn}
\usepackage{amsmath}
\usepackage{epsfig}

\long\def\inst#1{\par\nobreak\kern 4pt\nobreak
    {\it #1}\par\vskip 10pt plus 3pt minus 3pt}

\begin{document}
{\pagestyle{empty}
\begin{flushleft}
BABAR-CONF-07/035 
\end{flushleft}

\title{Heavy Quarkonium Spectroscopy}

\author{R. Faccini}
\affiliation{
University "La Sapienza" and INFN Rome, Dipartimento di 
Fisica, 2 P$_{\rm le}$ Aldo Moro, I-00185, Rome, Italy
}
\maketitle
} 
\section{introduction}
Although the Standard Model of elementary particles is well 
established, strong interactions are not yet fully under control. We 
believe QCD is the field theory capable of describing them, but we are 
not yet capable, in most of the cases, to make exact
predictions. Systems that include heavy quark-antiquark pairs 
(quarkonia) are ideal and unique laboratories to proble both the high 
energy regimes of QCD, where an expansion in terms of the coupling 
constant is possible, and the low energy regimes, where 
non-perturbative effects dominate.

In the last years this field is experiencing a rapid expansion with a 
wealth of new data coming in from diverse sources:
data on quarkonium formation from dedicated experiments (BES at 
BEPC, KEDR at VEPP-4M CLEO-c at CESR), clear samples produced by high 
luminosity B-factories (PEP and KEKB), and very large samples produced 
from  gluon-gluon fusion in $\rm{p\bar{p}}$ annihilations at Tevatron 
(CDF and D0 experiments).
\begin{figure}[bht]
\epsfig{file=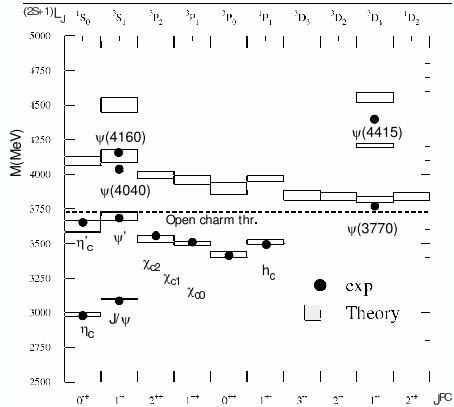,height=6cm}
 \caption{\it Charmonium states with 
$L<=2$. The theory predictions are according to the potential models 
described in Ref.~\cite{Brambilla:2004wf}. 
 }
    \label{fig:charmon}
\end{figure}

In this review I will first summarize recent developments in the 
understanding of heavy quarkonium states which have a well established 
quark content.
\begin{figure}[bht]
\epsfig{file=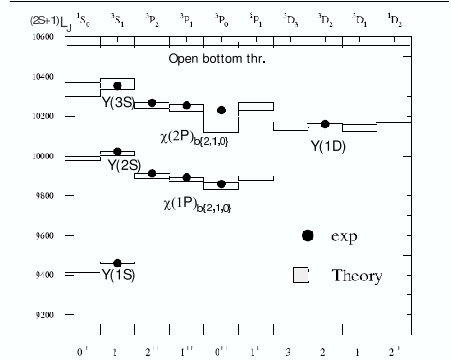,height=6cm}
 \caption{\it  Bottomonium (right) with 
$L<=2$. The theory predictions are according to the potential models 
described in Ref.~\cite{Brambilla:2004wf}. 
 }
    \label{fig:bottom}
\end{figure}

Next, the core of the paper will be spent to review the experimental 
evidences of new states that might be aggregations of more than just a 
quark-antiquark pair. Although the possibility to have bound states of 
two quarks and two antiquarks or of quark-antiquark pairs and gluons 
has been predicted since the very start of the quark 
model~\cite{GellMann:1964nj}, 
no observed state has yet been attributed to one of them: achieving 
such an attribution would be a major step in the understanding of 
the strong interactions. 
\begin{table*}[!htb]
\begin{center} 
\caption{ 
Most recent determination of the  $J^{PC}=1^{--}$  charmonium states from BES~\cite{Ablikim:2007gd}, compared to the 2006 edition of the PDG~\cite{Yao:2006px} 
}
\label{tab:rscan} 

\begin{tabular}{|c|c|c|c|c|c|} \hline \hline
        &         &$\psi(3770)$  &$\psi(4040)$&$\psi(4160)$&$\psi(4415)$\\ \hline
$M$     &PDG2006  &3771.1$\pm$2.4&4039$\pm$1.0&4153$\pm$3  & 4421$\pm$4 \\
(MeV/$c^2$)          &BES '07 &3771.4$\pm$1.8& 4038.5$\pm$4.6 & 4191.6$\pm$6.0 & 4415.2$\pm$7.5\\\hline\hline
$\Gamma_{tot}$&PDG2006  &23.0$\pm$2.7&80$\pm$10    &103$\pm$8    &     62$\pm$20   \\
(MeV)         &BES '07      &25.4$\pm$6.5&81.2$\pm$14.4&72.7$\pm$15.1&73.3$\pm$21.2\\\hline\hline
\end{tabular} 
\end{center}
\end{table*}

The currently most credited possible states beyond the mesons and the 
baryons are (you can find a review in~\cite{Brambilla:2004wf}):
\begin{itemize}
\item hybrids: bound states of a quark-antiquark pair and a number of 
gluons. The lowest lying state is expected to have quantum numbers 
$J^{PC}=0^{+-}$. The impossibility of a quarkonium state to assume 
these quantum numbers (see below) makes this a unique signature for 
hybrids. Alternatively a good signature would be the preference to 
decay into a quarkonium and a state that can be produced by the excited 
gluons (e.g. $\pi^+\pi^-$ pairs).
\item molecules: bound states of two mesons, usually represented as 
$[Q\bar{q}][q^{\prime}\bar{Q}]$, where $Q$ is the heavy quark. The 
system would be stable if the binding energy would set the mass of the 
states below the sum of the two meson masses.
While this could be the case for when $Q=b$, this does not apply for 
$Q=c$, where most of the current experimental data are. In this case 
the two mesons can be bound by pion  exchange. This means that only 
states decaying strongly into pions can bind with other mesons (e.g. 
there could be $D^*D$ states), and 
that the bound state could decay into it's constituents.
\item tetraquarks: a quark pair bound with an antiquark 
pair, usually represented as $[Qq][\bar{q^{\prime}}\bar{Q}]$. A full 
nonet of states is predicted for each spin-parity, i.e. a large amount 
of states is expected. There is no need for these states to be close to 
any threshold.
 \end{itemize}
In setting after these states one must also beware of threshold effects, 
where amplitudes might be enhanced when new hadronic final states become 
possible.

This paper will summarize the latest findings on the 
spectroscopy  of the known heavy quarkonium states and 
the status of the art of the understanding of all other states which 
might not fit in the ordinary spectroscopy.
\section{Heavy quarkonium spectroscopy}
The heavy quark inside these bound states has low 
enough energy that the corresponding
spectroscopy is close to the non-relativistic interpretations of the 
atoms. The quantum numbers that are more appropriate to characterize a 
state are therefore, in decreasing order of energy splitting among 
different eigenstates, the radial excitation ($n$), the spatial angular 
momentum $L$, the spin $S$ and the total angular momentum $J$. Given 
this set of quantum numbers, the parity and charge conjugation of the 
states are derived by $P=(-1)^{(L+1)}$ and $C=(-1)^{(L+S)}$. 
Figures~\ref{fig:charmon} and~\ref{fig:bottom} show the mass and 
quantum number assignments of 
the 
well established charmonium and bottomonium states.

\subsection{Charmonium spectroscopy}
Figure~\ref{fig:charmon} shows that all the predicted states below 
open charm
 threshold have been observed, leaving the search open only to states above the
 threshold. In this field the latest developments concern the measurement of 
the paramaters and the quantum number assignment for the $J^{PC}=1^{--}$ 
states.

The BES collaboration has recently performed a fit to the $R$ scan 
results which takes into account interference between resonances more 
accurately~\cite{Ablikim:2007gd}. The updated parameters are reported in Tab.~\ref{tab:rscan}, compared with the most recent determinations.

The $J^{PC}=1^{--}$ assignment does not univocously identify the state, 
since both $^{2S+1}L_J=^3D_1$ and $^3D_1$ states would match it. The 
recent observation from Belle of the first exclusive decay of the 
$\psi(4415)\to DD^*_{2}(2460)$~\cite{belle:2007fq}, shows that this 
meson is predominantly $D$ wave. At the same time the study from CLEO-c 
of the $\psi(3770)\to\chi_{cJ}\gamma$~\cite{Briere:2006ff} confirms the dominance of the $D$ wave also in this meson. Both these assignments confirm the theoretical predictions as shown in Fig.~\ref{fig:charmon}.

\subsection{Bottomonium transitions}
Figure~\ref{fig:bottom} shows that the panorama in the bottomonium 
sector is
 much less complete, since there is a large number of states below the open bottom 
threshold which have not yet been observed. Moreover there is no recent 
measurement on the topic.

\begin{figure}[htb]
\epsfig{file=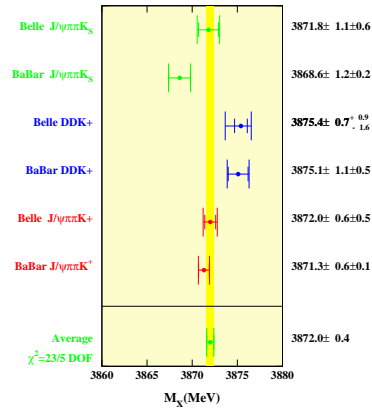,height=6cm} 
 \caption{\it Measured mass of the $X(3872)$ particle. The different 
production modes ($B^0\to XK_S$ and $B^-\to X K^-$)
and the different decay modes ($X\to J/\psi\pi\pi$ and $X\to 
D^{*0}D^0$) are separated.} 
\label{fig:xmass}
\end{figure}

There are, on the contrary, plenty of results on the transitions 
between $J^{PC}=1^{--}$ states, i.e. between $Y(nS)$ 
and $Y(mS)$ states. These transitions are relevant because it is possible to predict both the 
di-pion invariant mass spectrum in the case of $Y(nS)\to Y(mS)\pi^+\pi^-$ decays, and the relative rate between these 
decays and the $Y(nS)\to Y(mS)\eta$ decays. These transitions allow therefore stringent tests of low energy QCD, in particular
of the predicitons of the Multipole Expansion~\cite{Kuang:2006me}. 

Recent measurements of the $Y(4S)\to Y(2S)\pi^+\pi^-$ decays from BaBar~\cite{Aubert:2006bm} showed a discrepancy with 
the above-mentioned predictions. Since such predictions mutuated the matrix elements from PCAC, CLEO-c recently published a 
comprehensive study of $Y(3S)\to Y(mS)\pi^+\pi^-$ decays under more general assumpions~\cite{cleo:2007sja}: a fit to the two distributions ($m=1,2$)
letting the matrix elements float shows a good agreement with the data, thus confirming the validity of the model. 
\section{Non-standard charmonium states}
\subsection{The $X(3872)$}

The $X(3872)$ was the first state that was found not to easily fit charmonium spectroscopy.
 It was initially observed decaying into $J/\psi\pi^+\pi^-$ with a mass just beyond the open charm 
threshold~\cite{Choi:2003ue}. The $\pi^+\pi^-$ invariant mass distribution 
preferred the hypothesis of a 
$X(3872)\to J/\psi\rho$ decay, which would have indicated that if this 
were a charmonium state, the decay would 
have violated the isospin. Since it would be quite unusual to have the dominant decay to be isospin violating,
a search of the isospin partner $X^+\to J/\psi\rho$ was conducted 
invain by BaBar~\cite{Aubert:2004zr}.
In the meanwhile the decay $X\to J/\psi\gamma$ was 
observed~\cite{Aubert:2006aj}, implying
positive intrinsic charge conjugation.

The most recent developments concern the final assessment of the $J^{PC}$ of this particle and the indication
that the $X(3872)$ is a actually a doublet. The CDF collaboration has infact performed the full angular analysis
of the $X\to J/\psi\pi\pi$ decay~\cite{Abulencia:2006ma} concluding 
that $J^{PC}=1^{++}$ and $2^{-+}$ are 
the only assignments consistent with data. It also confirmed that the decays has a $\rho$ as intermediate state.
Combining this information with the preliminary result from Belle~\cite{Abe:2005iya} which rules out the $2^{-+}$ 
hypothesis, the only possible assignment is $J^{PC}=1^{++}$.
\begin{table*}[htb]
\begin{center} 
\caption{ 
Measured $X(3872)$ branching fractions, separated by production and decay mechanism. The ratio of the 
measurements in the two production mechanisms is also reported as $R_{0/+}=BF(B\to XK^-)/BF(B\to XK^0)$. A '$^*$' indicates numbers which are 
derived from the published values by assuming gaussian uncorrelated errors.
}
\label{tab:xsec} 
\begin{tabular}{|l|c|c||c|} \hline 
     & BaBar & Belle & combined\\ \hline
BF($B\to XK^-$)BF($X\to J/\psi\pi\pi$)$\times 10^5$&1.01$\pm0.25\pm0.10$~\cite{Aubert:2005zh}&1.05$\pm$0.18~\cite{Choi:2003ue}&1.04$\pm0.15^*$\\   
BF($B\to XK^0$)BF($X\to J/\psi\pi\pi$)$\times 10^5$&0.51$\pm0.28\pm0.07$~\cite{Aubert:2005zh}&$0.99\pm0.33^*$&$0.72\pm0.22^*$\\   
BF($B\to XK^-$)BF($X\to D^{*0}D^0$)$\times 10^5$   
&$17\pm4\pm6$~\cite{babar:2007rv}&10.7$\pm 3.1^{+1.9}_{-3.3}$~\cite{Gokhroo:2006bt}&$12\pm4^*$\\   
BF($B\to XK^0$)BF($X\to D^{*0}D^0$)$\times 10^5$    
&$22\pm10\pm5$~\cite{babar:2007rv}&17$\pm7^{+3}_{-5}$~\cite{Gokhroo:2006bt}&$18\pm7^*$\\  
\hline
$R_{0/+}$ with $X\to J/\psi\pi\pi$            	   
&$0.50\pm0.30$~\cite{Aubert:2005zh}&$0.94\pm0.26$&$0.75\pm0.20^*$\\ 
$R_{0/+}$ with $X\to D^{*0}D^0$            	   
&$1.3\pm0.7$~\cite{babar:2007rv}&$1.6\pm0.6^*$&$1.5\pm0.5^*$\\ \hline
\end{tabular} 
\end{center}
\end{table*}

 As far as the mass and width of the $X(3872)$ are concerned, BaBar has 
published an analysis of the $B\to XK$
 decays with $X\to D^{*0}D^0$~\cite{babar:2007rv} while Belle has 
updated the mass measurements in $X\to J/\psi\pi\pi$
 decays~\cite{bellex}. The summary of all available mass measurements 
is shown in Fig.~\ref{fig:xmass} where 
 the measurements are separated by production and decay channel. There 
is an indication that the particle
 decaying into $J/\psi\pi\pi$ is different from the one decaying into 
$D^{*0}D^0$, their masses differing
by about 4 standard deviations.

In addition, the BaBar paper contains also a first measurement of the $X(3872)$ width,
$\Gamma=$($3.0^{+4.6}_{-2.3}\pm 0.9$)MeV. Finally the measurements of $X$ the branching 
fractions in $J/\psi\pi\pi$ and $D^{*0}D^0$ are summarized in Tab.~\ref{tab:xsec}.

\subsection{The $1^{--}$ family}
The easiest way to assign a value for $J^{PC}$ to a particle is to observe its production via $e^+e^-$ 
annihilation, where the quantum numbers must be the same as the the photon:  $J^{PC}=1^{--}$. $B$ factories
can investigate a large range of masses for such particles by looking for events where the initial state 
radiation brings the $e^+e^-$ center-of-mass energy down to the particle's mass (the so-called 'ISR' events). 
Alternatively, 
dedicated $e^+e^-$ machines, like CESR and $BEP$ scan directly the certer-of-mass energies of interest.
\begin{figure}[hbt]
\epsfig{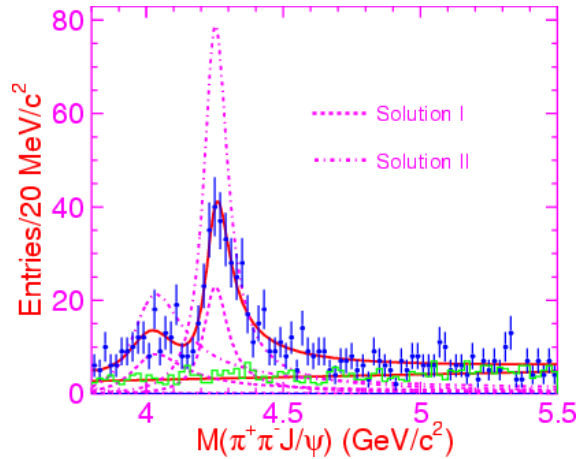}
 \caption{\it $J/\psi\pi^+\pi^-$  invariant mass in ISR production.}
\label{fig:belle1mm}
\end{figure}
\begin{figure}[tbh]
\epsfig{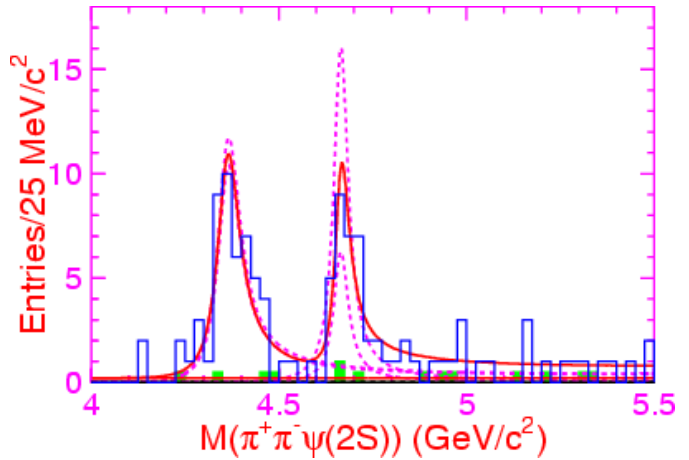}
 \caption{\it  $\psi(2S)\pi^+\pi^-$ invariant 
mass in ISR production.}
\label{fig:belle1mm2S}
\end{figure}

The observation of new states in these processes started with the discovery of the $Y(4260)\to 
J/\psi\pi^+\pi^-$ 
by BaBar~\cite{Aubert:2005rm}, promptly confirmed both in the same production process~\cite{He:2006kg} and in 
direct production by CLEO-c~\cite{Coan:2006rv}. The latter paper also reported evidence for 
$Y(4260)\to J/\psi\pi^0\pi^0$ and some events of $Y(4260)\to J/\psi K^+K^-$. 

While investigating whether the $Y(4260)$ decayed to 
$\psi(2S)\pi^+\pi^-$ BaBar found that such decay did not exist but 
discovered a new $1^{--}$ state, the $Y(4350)$~\cite{Aubert:2006ge}. 
While the absence of $Y(4260)\to \psi(2S)\pi^+\pi^-$ decays could be 
explained if the pion pair in the $J/\psi\pi^+\pi^-$ decay were 
produced 
with an intermediate state that is to amssive to be produced with a 
$\psi(2S)$ (e.g. an $f^0$), the absence of $Y(4350)\to 
J/\psi\pi^+\pi^-$ is still to be understood, more statistics might be 
needed in case the $Y(4260)$ decay hides the $Y(4350)$.

Recently Belle has published the confirmation of all these $1^{--}$ 
states~\cite{belle:2007sj,belle:2007ea} and at the same time has 
unveiled a new states that was not visible in BaBar data due to the 
limited statistics: the $Y(4660)$. Figures~\ref{fig:belle1mm} 
and~\ref{fig:belle1mm2S} show the 
published invarianet mass spectra for both the $J/\psi\pi^+\pi^-$ and 
the $\psi(2S)\pi^+\pi^-$ decays.

\begin{figure*}[bht]
\epsfig{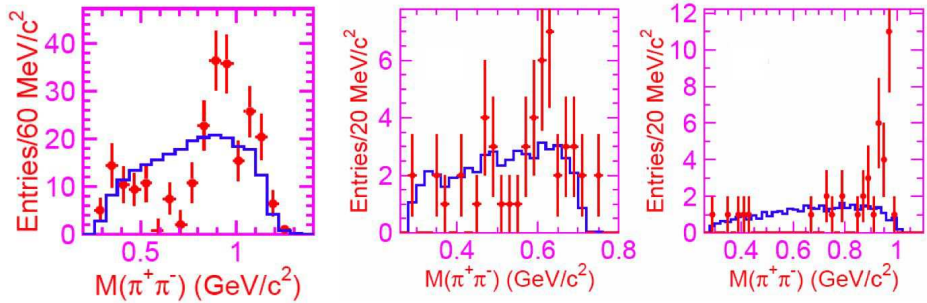}
 \caption{\it Di-pion invariant mass distribution in $Y(4260)\to 
J/\psi \pi^+\pi^-$ (left), $Y(4350)\to \psi(2S) \pi^+\pi^-$ (center), 
and $Y(4660)\to \psi(2S) \pi^+\pi^-$ (right) decays. 
}
\label{fig:bellepipiinv}
\end{figure*}

A critical information for the unravelling of the puzzle is whether the 
pion pair comes from a resonant state. Figure~/ref{fig:bellepipiinv} 
shows the di-pion invariant mass spectra published by Belle for all the 
regions where new resonances have been observed. Although the 
subtraction of the continuum is missing, there is some
indication that only the $Y(4660)$ has a well defined intermediate 
state (most likely an $f_0$, while others have a more complex 
structure.

A discriminant measurement between Charmonium states and new 
aggregation forms is the relative decay rate between these decays into 
Charmonium and the decays into two charm mesons. Searches have 
therefore been carried out for $Y\to D^{(*)}D^{(*)}$ 
decays~\cite{Abe:2006fj,collaboration:2007mb,Aubert:2007pa} without any 
evidence for a signal. The most stringent limit is~\cite{Aubert:2007pa} 
$BF(Y(4260)\to D\bar{D})/BF(Y(4260)\to J/\psi\pi^+\pi^-)<1.0 @$ 90\% 
confidence level.
\begin{figure}[hbt]
\epsfig{file=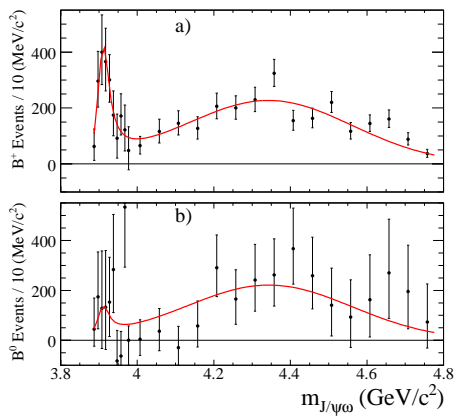,height=6cm}
 \caption{\it The
$J/\psi \omega$ distribution in a) $B\to J/\psi \omega K^+$ and b)  
$B\to J/\psi \omega K_S$ decays. The superimposed line is the result of 
the fit to the data.
\label{fig:babarjpsiomega}}
\end{figure}

\subsection{The $3940$ family}
Three different states have been observed in the past years by 
the Belle collaboration with masses close to $3940 \rm {Mev/c}^2$: one, 
named 
$X$, observed in continuum events 
(i.e. not in $Y(4S)$ decays) produced in pair with a $J/\psi$ 
meson and decaying into $DD^{*}$~\cite{Abe:2007jn}; a second one, named 
 $Y$, observed in $B$ decays and decaying into 
$J/\psi\omega$~\cite{Abe:2004zs}; a third 
one, named $Z$ produced in two-photon reactions and decaying into 
$D$-pairs~\cite{Uehara:2005qd}. While the $X$ is consistent with both 
$J^{PC}=0^{+-}$ and $1^{++}$, the quantum number assignment of the $Y$ 
and the $Z$ states is clear: $J^{PC}=1^{++}$ and $2^{++}$ respectively.
Finally the $Y$ is the only apparently broad state 
($\Gamma=87\pm34$MeV).

Because of these quantum number assignments and their masses these 
states are good candidates for the radial eccitation of the $\chi$ 
mesons, in particular the $Z(3940)$ meson could be identified with the 
$\chi_{c0}(2P)$ and the $Y(3940)$ with the $\chi_{c1}(2P)$.
The unclear points are the identification of the $X(3940)$ state and 
the explanation of why the $Y(3940$ state does not decay preferentially 
in $D$ mesons.

The most recent develpment on this topic is the confirmation from 
the BaBar collaboration of the $Y(3940)\to J/\psi\omega$ 
decays~\cite{Aubert:2007vj}. The analysis utilizes the decay properties 
of the $\omega$ meson to extract a clean signal (see 
Fig.~\ref{fig:babarjpsiomega}).
The interesting part is that he mass and 
the width measured in this paper are lower than when observed, albeit 
consitent 
($m_Y=3914.6^{+3.8}_{-3.4}(stat.)\pm1.9(sys.)$Mev$/c^2$, 
$\Gamma_Y=33^{+12}_{-8}(stat.)\pm5(sys.)$MeV), opens the interesting 
possibility that the $X$ and the $Y$ particles be the same, thus 
solving the two abovementioned open issues.

\subsection{The $X(4160)$}
As we have already discussed, it is critical to investigate decay 
channels of the new states into $D$ meson pairs. Unfortunately the 
detection efficiency for $D$ mesons in low, due to the large number of 
possible decays. The Belle collaboration has developed a partial 
reconstruction technique 
that allows to overcome this limitation in the case of new states  
produced in continuum paired with known Charmonium 
states~\cite{belle:2007sy}. The Charmonium is fully reconstructed, 
while only one of the two D mesons is reconstructed. The kinematics of 
the other is inferred from the known center-of-mass energy and the 
different possibile $D$ mesons are discriminated on the basis of the 
missing mass.

This technique has allowed the confirmation of the $X(3940)$ production 
and decay, and, most interestingly, the observation of 
the $X(4160)$ state, decaying into $D D^{*}$. Given the fact that, for 
reasons yet to be understood, continuum events seem to produce ont 
$J^{PC}=0^{-+}$ or $1^{++}$ states in pair with the $J/\psi$ and since 
the measured mass is consistent with the expectations of a radial 
excitation of the $\eta_c$, this new state is likely to be an 
$\eta_c(3S)$.
\begin{figure}[tbh]
\epsfig{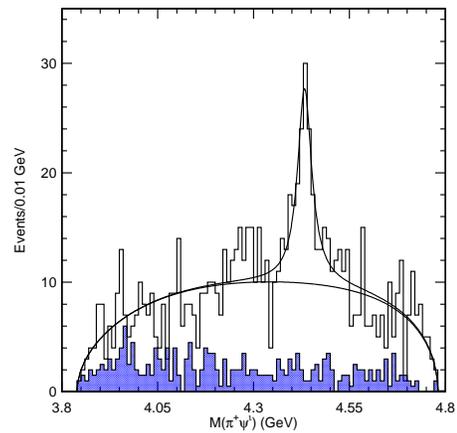}
 \caption{\it The $\psi(2S)\pi$ invariant mass distribution in 
$B\to \psi(2S)\pi K$ decays.}
\label{fig:belleZ4430}
\end{figure}

\subsection{The first charged state: $Z(4430)$}
The real turning point in the query for states beyond the Charmonium 
was the observation by the Belle Collaboration of a charged state
decaying into $\psi(2S)\pi^\pm$~\cite{belle:2007wg}. 
Figure~\ref{fig:belleZ4430}
shows the fit to the  $\psi(2S)\pi$ invariant mass distribution in
$B\to \psi(2S)\pi K$ decays, returning a mass $M=4433\pm4\rm{MeV/c}^2$  
and a width $\Gamma=44^{+17}_{-13}$ MeV. Due to the relevance of such 
an observation a large number of tests has been performed, breaking the 
sample in several subsamples and finding consistent results in all 
cases. Also, the possibility of a reflection of a $B\to\psi(2S)K^**$  
decay has been falsified by explicitely vetoing windows in  the $K\pi$ 
invariant mass.

In terms of quarks, such a state must 
contain a $c$ and a $\bar{c}$, but given its charge 
it must also contain at least a $u$ and a $\bar{d}$. The only open 
options are the tetraquark, the molecule or the threshold effects. The 
latter two options are possible due to the closeness of the $D_1D^*$ 
threshold. 
\begin{figure}[htb]
\epsfig{file=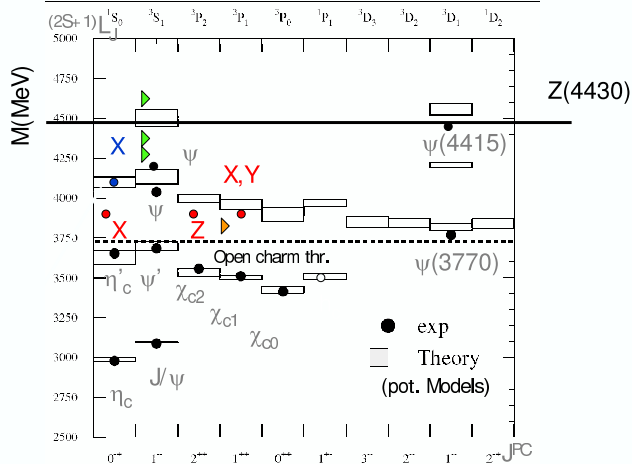, height=6cm} 
 \caption{\it Measured masses of the newly observed states, positioned 
in the spectroscopy according 
to their most likely quantum numbers. The charged state ($Z(4430)$) has 
clearly no $C$ quantum number.}
\label{fig:summary}
\end{figure}

Finding the corresponding neutral state, observing a decay mode of the 
same state or at least having a confirmation of its existence, are 
critical before a complete picture can be drawn.

\section{Conclusions}
More than 30 years after its first observation, the heavy-quarkonium is 
still 
a valid test ground for understanding QCD. The study of well 
established quarkonium states 
yields information on low energy QCD while the undestanding of the 
quarkonium spectroscopy, 
predictable in potential models, allows searches for different 
aggregation states than the 
long established mesons.

The high statistics and quality data from B-Factories have produced a 
very large number of new
 states whose interpretation is still a matter of debate. This paper 
attempted a categorized 
review of all these states. A full summary, including the most likely 
quantum number assignment 
is shown in Fig.~\ref{fig:summary}. Lots of theoretical models have 
been developed to interpret 
the situation but the picture is far from complete: more precise 
predictions are needed from theory
and a systematic experimental exploration of all possibile production 
and decay mechanisms of these new states 
is still in the works. 
\section{Acknowledments}
I would like to thank for the help I received in preparing this talk 
from my collegues in the BaBar, Belle, Cleo, BES, CDF and D0  
collaborations. I would also like to thank Luciano Maiani and Antonello 
Polosa for the continous discussion on the topic.

\end{document}